\documentclass{emulateapj} 
\usepackage{epsfig}

\newcommand{\be}{\begin{equation}}
\newcommand{\ee}{\end{equation}}
\newcommand{\bea}{\begin{eqnarray}}
\newcommand{\eea}{\end{eqnarray}}

\usepackage{amsmath}
\usepackage{graphicx}

\usepackage[usenames]{color}

\begin{document}

\author{Louis E. Strigari}
\affil{Mitchell Institute for Fundamental Physics and Astronomy, 
Department of Physics and Astronomy, Texas A \& M University, 
College Station, TX 77843, USA}
\author{Carlos S. Frenk}
\affil{Institute for Computational Cosmology, Dep. of Physics,
    University of Durham, South Road, Durham  DH1 3LE, UK} 
\author{Simon D. M. White}
\affil{Max-Planck-Institut f\"{u}r Astrophysik,
Karl-Schwarzschild-Stra\ss{}e 1, 85740 Garching bei M\"{u}nchen,
Germany}

\title{Dynamical constraints on the dark matter distribution of the Sculptor dwarf spheroidal from stellar proper motions}

\begin{abstract}
We compare the transverse velocity dispersions recently measured
within the Sculptor dwarf spheroidal galaxy to the predictions of
previously published dynamical models. These provide good fits to the
observed number count and velocity dispersion profiles of metal-rich
and metal-poor stars both in cored and in cusped potentials. At the
projected radius where the proper motions were measured, these models
predict transverse dispersions in the range 6 to 9.5 km/s, with the
tangential dispersion about 1 km/s larger than the (projected) radial
dispersion. Both dispersions are predicted to be about 1 km/s larger
for metal-poor than for metal-rich stars. At this projected radius,
cored and cusped potentials predict almost identical transverse
dispersions. The measured tangential dispersion ($8.5 \pm 3.2$ km/s)
agrees remarkably well with these predictions, while the measured
radial dispersion ($11.5 \pm 4.3$ km/s) differs only at about the
$1\sigma$ level. Thus, the proper motion data are in excellent
agreement with previous data but do not help to distinguish between
cored and cusped potentials. This will require velocity dispersion
data (either from proper motions or from radial velocities) with
uncertainties well below 1 km/s over a range of projected radii.

\end{abstract}

\maketitle

\section{Introduction} 
\par The kinematics of stars in the central regions of dark
matter-dominated dwarf spheroidal galaxies (dSphs) may allow us to
distinguish between different hypotheses for the nature of dark
matter. On these scales the simplest ``Lambda cold dark matter''
($\Lambda$CDM) model of structure formation predicts cuspy halos that
are well described by the NFW
profile~\citep{Navarro:1995iw,NFW:1997, 2010MNRAS.402...21N}.
Alternative forms of dark matter, such as
self-interacting~\citep{Spergel:1999mh,Vogelsberger:2012ku,Rocha:2012jg} or
warm~\citep{Bode:2000gq,Lovell:2015psz} dark matter predict low-mass
haloes with either a constant
density core or a shallower central cusp.

Published analyses of the line-of-sight (LOS) velocities of stars in
dSphs have led to contradictory conclusions: some authors conclude that the
kinematic data require dark matter
cores~\citep{Gilmore:2007fy,Agnello:2012uc,Walker:2011zu}, while
others find that the data are also consistent with NFW-like
cusps~\citep{Strigari:2010un,Breddels:2013,Richardson:2013lja,Strigari:2014yea}.

\par Accurate measurements of stellar proper motions (PMs) can further
constrain the central density in
dSphs~\citep{Wilkinson:2001ut,Strigari:2007vn,2017MNRAS.471.4541R}.
Together with LOS velocities and positions on the sky, the two
additional transverse velocity components provide five out of the six
phase space coordinates of the stars. These data can help break the
degeneracy between stellar velocity anisotropy and galaxy mass
profile, which has so far been a limiting factor in kinematical
studies of dSphs.  Since thousands of LOS velocities are already
available for the best studied cases, such as Fornax and
Sculptor~\citep{Walker:2008ax}, precise measurements of the transverse
velocity dispersions are required to improve constraints on the inner
dark matter density profile.
 
\par In a recent landmark study~\citet{2017arXiv171108945M} have
measured sufficiently accurate proper motions for a number of stars in
Sculptor for the uncertainty in their transverse velocity dispersion
estimates to be significantly smaller than the intrinsic velocity
dispersion of the system. This is the first time such a measurement
has been performed in a dSph with this level of accuracy, thus finally
opening the era of using dynamical models with full phase space
information to constrain the inner dark matter distribution of dSphs.

\par Here, we use the self-consistent stellar distribution function
models for Sculptor published by~\citet{Strigari:2014yea} to predict
transverse velocity dispersions at the position where these were
measured by~\citet{2017arXiv171108945M}. Sculptor has two distinct
metallicity populations \citep{Battaglia:2008jz,Walker:2008ax}. The
\citet{Strigari:2014yea} models treat the metal-rich and metal-poor
components as separate populations orbiting in a common gravitational
potential which may be cusped or cored. They were fit directly to the
position and LOS velocity data of~\citet{Walker:2008ax}. As we will
see they predict transverse dispersions in good agreement with the new
measurements by~\citet{2017arXiv171108945M} in both cases. At the end of our 
paper we discuss the accuracy
of PM and LOS velocity dispersion measurements that will be required
to robustly distinguish a core from a cusp in the dark matter
distribution of this galaxy.
 
\section{Data}
\label{sec:data} 
\par To measure the PMs of stars in Sculptor
\citet{2017arXiv171108945M} combined astrometry from the first {\it
  Gaia} data release with positions from two overlapping HST fields,
which have an average projected radius $R = 185$~pc from the
center of Sculptor. With an approximate $12$ year baseline between the
observations, 15 stars with particularly well measured PMs allowed
them to estimate the transverse radial and tangential velocity
dispersions at this position in Sculptor as
$\sigma_R(R=185~\textrm{pc}) = 11.5 \pm 4.3$ km/s and
$\sigma_T(R=185~\textrm{pc}) = 8.5 \pm 3.2$ km/s, respectively.

\par The combination of these new transverse dispersion estimates with
earlier estimates of the LOS velocity dispersion profile of Sculptor
based on large numbers of stars~\citep{Battaglia:2008jz,Walker:2008ax}
provides the highest quality kinematic dataset ever assembled for a dSph.
There is some ambiguity in how best to identify the stars that belong
to the metal-rich (MR) and metal-poor (MP) populations and thus to
estimate the velocity dispersion and star count profiles for each
population~\citep{Battaglia:2008jz,Walker:2011zu,Amorisco:2011hb,Strigari:2014yea}.
Here we split the data of~\cite{Walker:2008ax} into distinct
populations as described in~\citet{Strigari:2014yea}; this results in
a sample of 397 MR and 763 MP stars.

\section{Models and data analysis} 
\label{sec:models}
\par We now briefly describe the dynamical model that we use to
analyse the Sculptor data and our statistical method for predicting the
transverse velocity dispersion from the~\cite{Walker:2008ax} data.

\par For the dark matter profile we consider two well-studied models,
one characterized by a central cusp and the other by a central
core. For the cusped model, we adopt the NFW profile
\citep{Navarro:1995iw,NFW:1997}, 
\be \rho(r) =
\frac{\rho_s}{x(1+x)^2},
\label{eq:NFW}
\ee  where 
$\rho_s$ is the scale density and $x = r/r_s$, with $r_s$  the scale radius.  For the
cored model we adopt the Burkert profile~\citep{Burkert:1995yz}, 
\be
\rho(r) = \frac{\rho_b}{(1+x_b)(1+x_b^2)},
\label{eq:burk}
\ee
where $\rho_b$ is the Burkert scale density and $x_b = r/r_b$, with $r_b$  the
Burkert scale radius.  

\par For the stellar distribution function we adopt the model
of~\citet{Strigari:2014yea}. Here we briefly review the relevant
aspects of this model and refer to this paper and to~\citet{White1981}
for a more detailed discussion.  The specific energy and specific
angular momentum of a star are $E = v^2/2 + \Phi(r)$ and $J = v r \sin
\theta$, respectively, where $v$ is the modulus of the velocity
vector, $\Phi(r)$ is the gravitational potential derived from the
density profile (assuming spherical symmetry), and $\theta$ is the
angle between this vector and the star's position vector relative to
the center of the system. If we assume that the dependence on $E$ and
$J$ is separable, we then have, \be f(E,J)=g(J)h(E),
\label{eq:df}
\ee
where both $g(J)$ and $h(E)$ are positive definite and have the simple
parametric forms as discussed below.

The stellar density profile and the radial and tangential velocity
dispersion profiles are given by
\bea \rho_\star (r) &=& 2 \pi \int_0^\pi
d \theta \sin \theta \int_0^{v_{esc}} dv v^2 g(J)h(E) \label{eq:rhostar}
\\ \rho_\star \sigma_r^2 (r) &=& 2 \pi \int_0^\pi d \theta \cos^2
\theta \sin \theta \int_0^{v_{esc}} dv v^4 g(J)h(E) \label{eq:sr}
\\ \rho_\star \sigma_t^2 (r) &=& \pi \int_0^\pi d \theta \sin^2 \theta
\sin \theta \int_0^{v_{esc}} dv v^4 g(J)h(E)
                       \label{eq:st}
\eea
where $v_{esc} = \sqrt{ 2 [\Phi_{lim} - \Phi(r) ]}$, with $\Phi_{lim}$ defined as the
value of the potential at the limiting radius. The total velocity dispersion at 
radius $r$ is then 
\be
\sigma_{\rm tot}^2(r) = \sigma_r^2(r) + 2\sigma_t^2(r).
\ee
 
\par Equations~\ref{eq:rhostar},~\ref{eq:sr}, and~\ref{eq:st} can be
combined to give the projected stellar density profile and the stellar
LOS and transverse velocity dispersion profiles at a fixed projected
distance, $R$:
\bea 
I_\star(R) &=& 2 \int_0^{\infty}  \rho_\star (r) dz,
\label{eq:IR} \\
I_\star(R) \sigma_{los}^2 (R) &=& 2 \int_0^{\infty}
\rho_\star(r)\frac{ z^2\sigma_r^2 +  R^2\sigma_t^2}{z^2+R^2} dz \\
I_\star(R) \sigma_R^2 (R) &=& 2 \int_0^{\infty}
\rho_\star(r)\frac{ R^2\sigma_r^2 +  z^2\sigma_t^2}{z^2+R^2} dz \\
I_\star(R) \sigma_T^2 (R) &=& 2 \int_0^{\infty}
\rho_\star(r) \sigma_t^2 dz 
\label{eq:sigmalos} 
\eea
where $r^2 = z^2 + R^2$. Here $R$ is the radial component in the plane of 
the sky, pointing radially outward from the center of the galaxy.  
The component $T$ is tangential to $R$, and is also in the plane of the sky. 

\par In the model of~\citet{Strigari:2014yea} the energy part, $h(E)$,
is a function of seven parameters; the angular momentum component,
$g(J)$, is a function of four parameters; and the density profiles in
Equations~\ref{eq:NFW} and ~\ref{eq:burk} are functions of two
parameters. Thus, for a single stellar population the model has 13
parameters. When performing a joint fit to both metallicity
populations, in which each population has independent distribution
function parameters but the common underlying dark matter profile is
still a function of two parameters, the model has a total of 24 free
parameters.

\par To fit this distribution function to the LOS velocities and the
photometry of~\citet{Walker:2008ax} for each population, we define a
likelihood function as in~\citet{Strigari:2014yea}.  From this
likelihood function we employ a Nested Sampling algorithm to scan
over the parameter space~\citep{Handley:2015fda,2015MNRAS.453.4384H}
and determine the posterior probability density for the model
parameters. The Nested Sampling algorithm is particularly useful for
sampling the posteriors of high dimensional parameter spaces which may
contain multiple regions of high probability.  In our code we use 500
live points and obtain the posterior probability densities for the
model parameters, as well as for parameters that are derived from
them. While there are degeneracies in the model parameters that
describe the stellar distribution function, the projected transverse
velocity dispersions, $\sigma_R$ and $\sigma_T$, which are derived from the
models, are well determined in all cases.

\section{Results} 

\par
\begin{figure*}
\begin{center}
\begin{tabular}{cc}
{\resizebox{6.0cm}{!}{\includegraphics[angle=0]{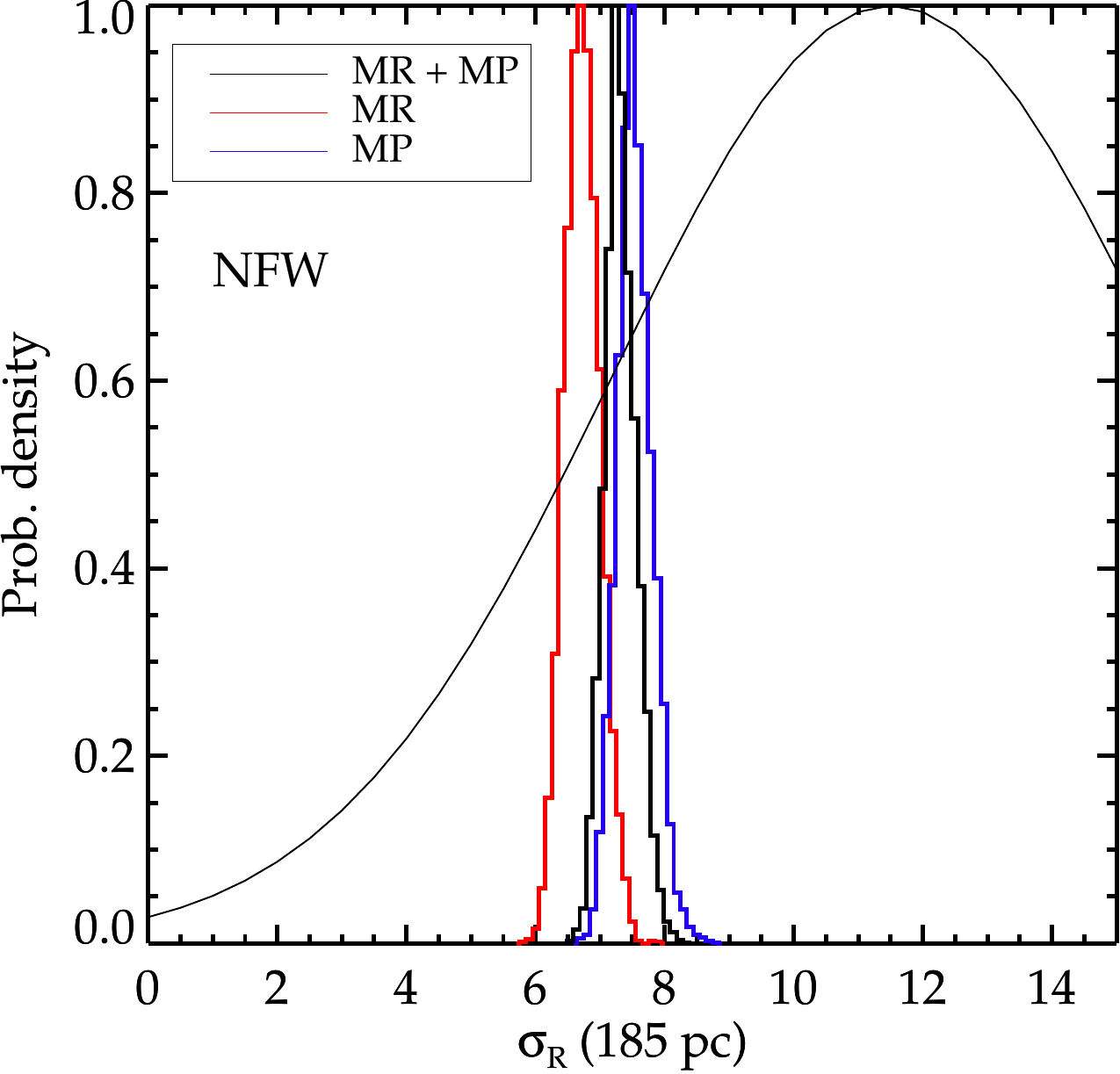}}} & 
{\resizebox{6.0cm}{!}{\includegraphics[angle=0]{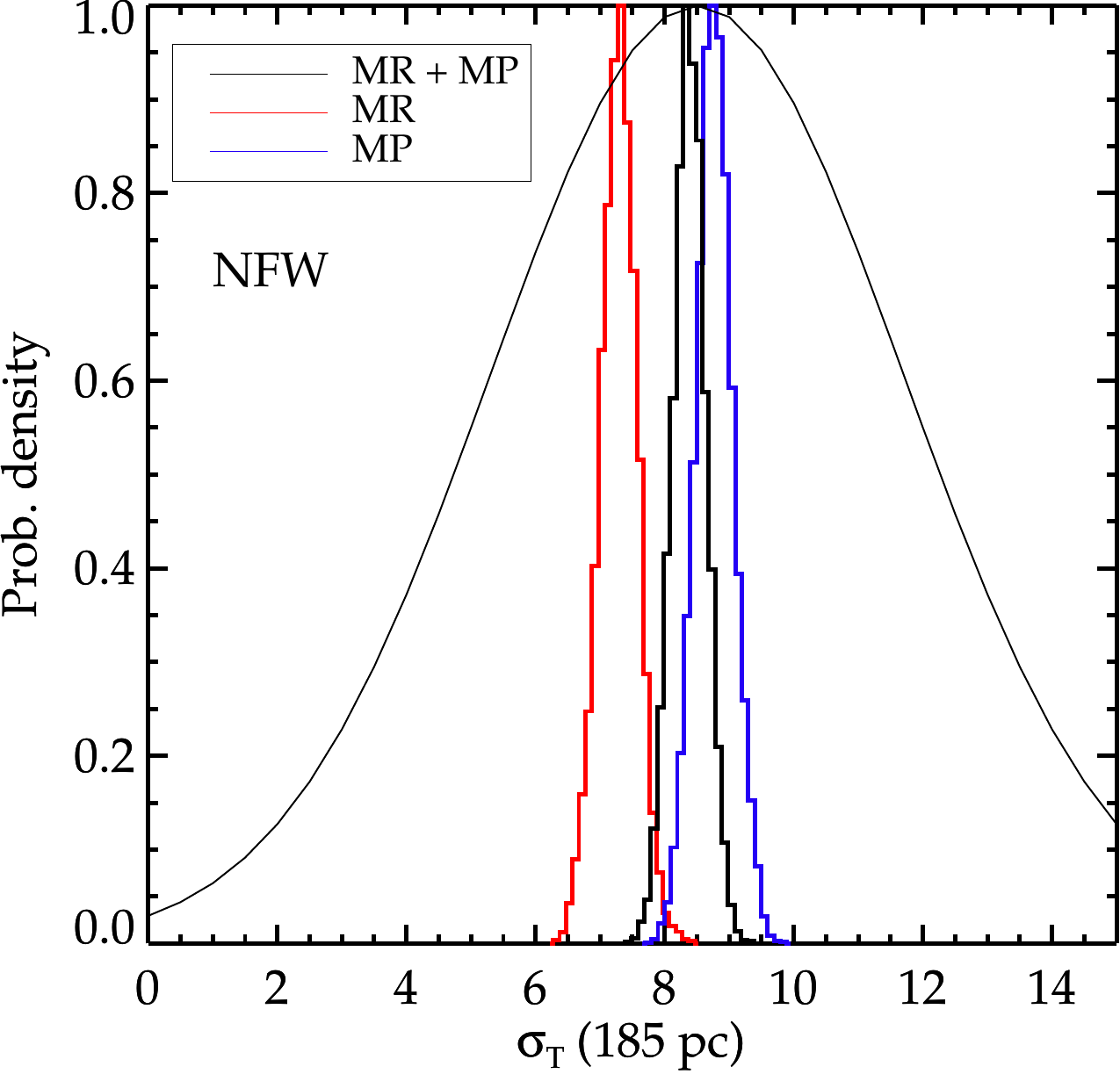}}} \\
{\resizebox{6.0cm}{!}{\includegraphics[angle=0]{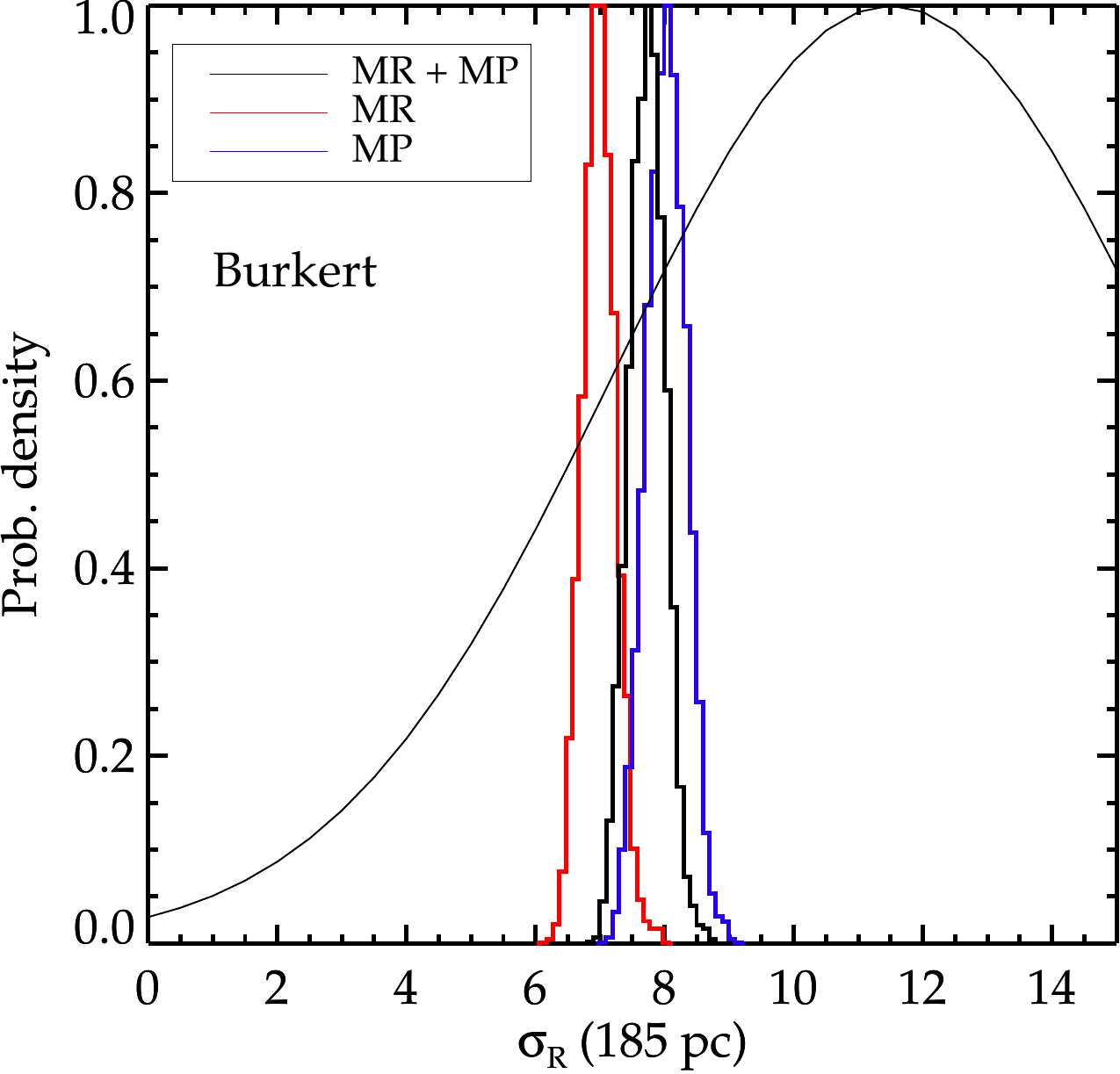}}} & 
{\resizebox{6.0cm}{!}{\includegraphics[angle=0]{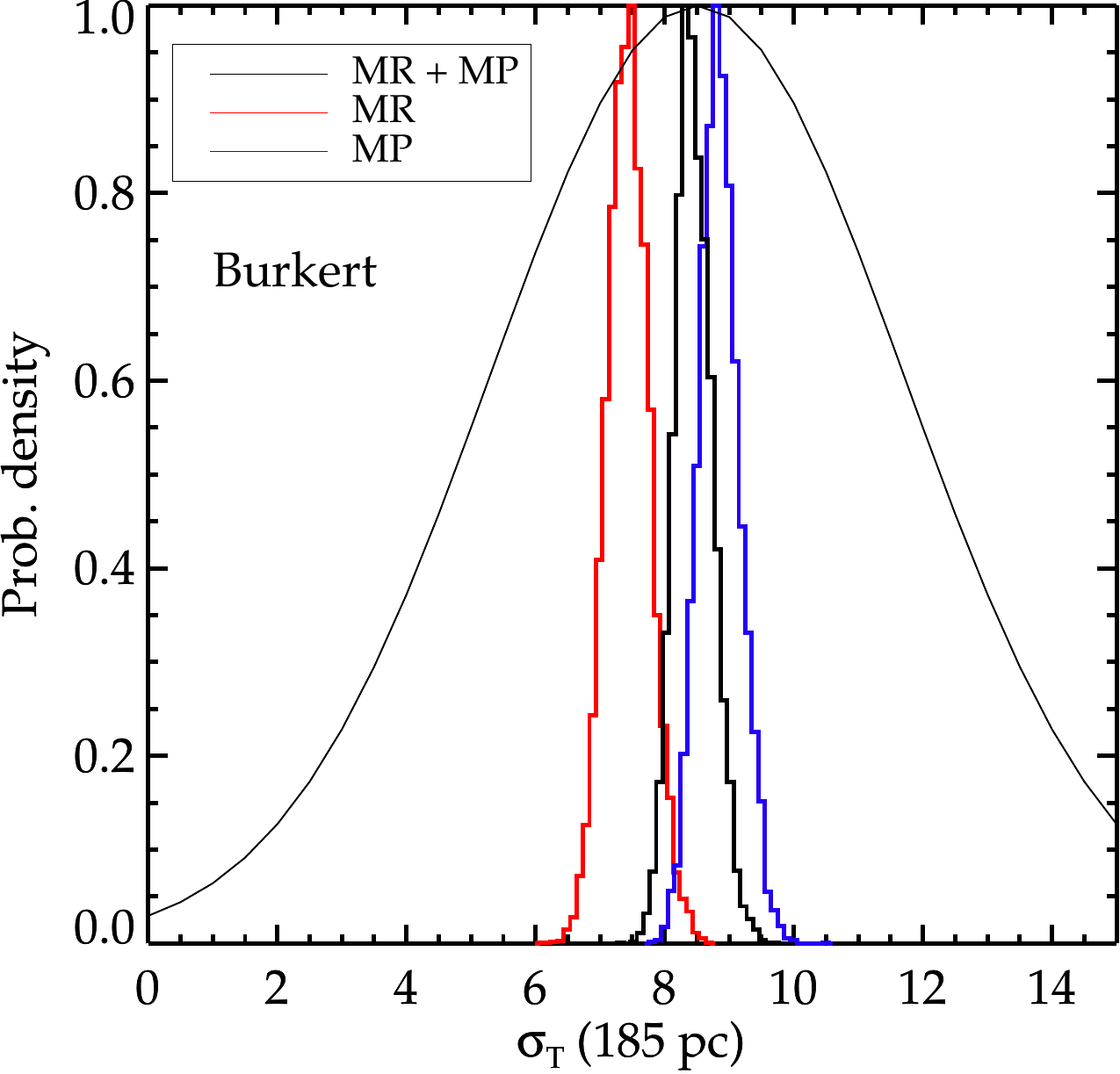}}} \\ 
\end{tabular}
\end{center}
\caption{Posterior probability densities for the transverse radial and
  tangential velocity dispersions for different populations, based on
  the models of \citet{Strigari:2014yea} which fit parametrised
  distribution functions to star count and line-of-sight velocity
  dispersion data from \citet{Walker:2011zu}.  The top row assumes an
  NFW dark matter profile and the bottom row a Burkert profile. The
  red histogram is for the metal rich population, the blue histogram
  for the metal poor population, and the black histogram for the total
  population. The thin black curves are Gaussian distributions that
  represent the PM-based estimates of \citet{2017arXiv171108945M} and
  their 1$\sigma$ uncertainty. }
\label{fig:mrmpmp}
\end{figure*}

 We now present predictions for the transverse velocity
dispersions at the position of the~\citet{2017arXiv171108945M}
measurements based on a joint analysis of the LOS velocity dispersion
and photometric count profiles for the MR and MP populations. The
posterior probability distributions for these transverse velocity
dispersions are shown in Figure~\ref{fig:mrmpmp} for three different
populations: the MP population, the MR population, and the total
population.  For this third case, the total dispersion at projected
radius, $R$, is:
\begin{equation} 
\sigma^2 (R) = f_{MR}(R) \sigma_{MR}^2(R) + [1-f_{MR}(R)] \sigma_{MP}^2(R), 
\label{eq:random}
\end{equation} 
where $f_{MR}(R)$ is defined as the fraction of MR stars at projected
radius, R. We evaluate $f_{MR}(R = 185 \, \textrm{pc})$ for each
photometric model, with the profiles normalized by the number of stars
in each population, accounting for the selection function
from~\citet{Walker:2011zu}. The fraction of MR stars is then given by
\begin{equation} 
f_{MR}(R) = \frac{I_{MR}(R)}{I_{MR}(R) + I_{MP}(R)},
\end{equation} 
where the $I$'s are the normalized photometric profiles. 

\par Figure~\ref{fig:mrmpmp} shows that there is excellent agreement
between the value of $\sigma_T$ predicted by our model based on the
~\cite{Walker:2008ax} data and that measured from the PMs. For
$\sigma_R$, the model prediction differs from the measurement at the
$\sim 1 \sigma$ level.  This is true for both NFW and Burkert
profiles. The model prediction agrees slightly better with the
measurement if we assume that the stars with measured PMs are randomly
drawn from the MP population or from the entire population (as
expressed in Equation~\ref{eq:random}) than if we assume they are
drawn from the MR population.  However, \citet{2017arXiv171108945M}
suggest, based on estimates of metallicity, that their PM sample may
be dominated by MR stars.

\par
\begin{figure*}
\begin{center}
\begin{tabular}{ccc}
{\resizebox{6.0cm}{!}{\includegraphics[angle=0]{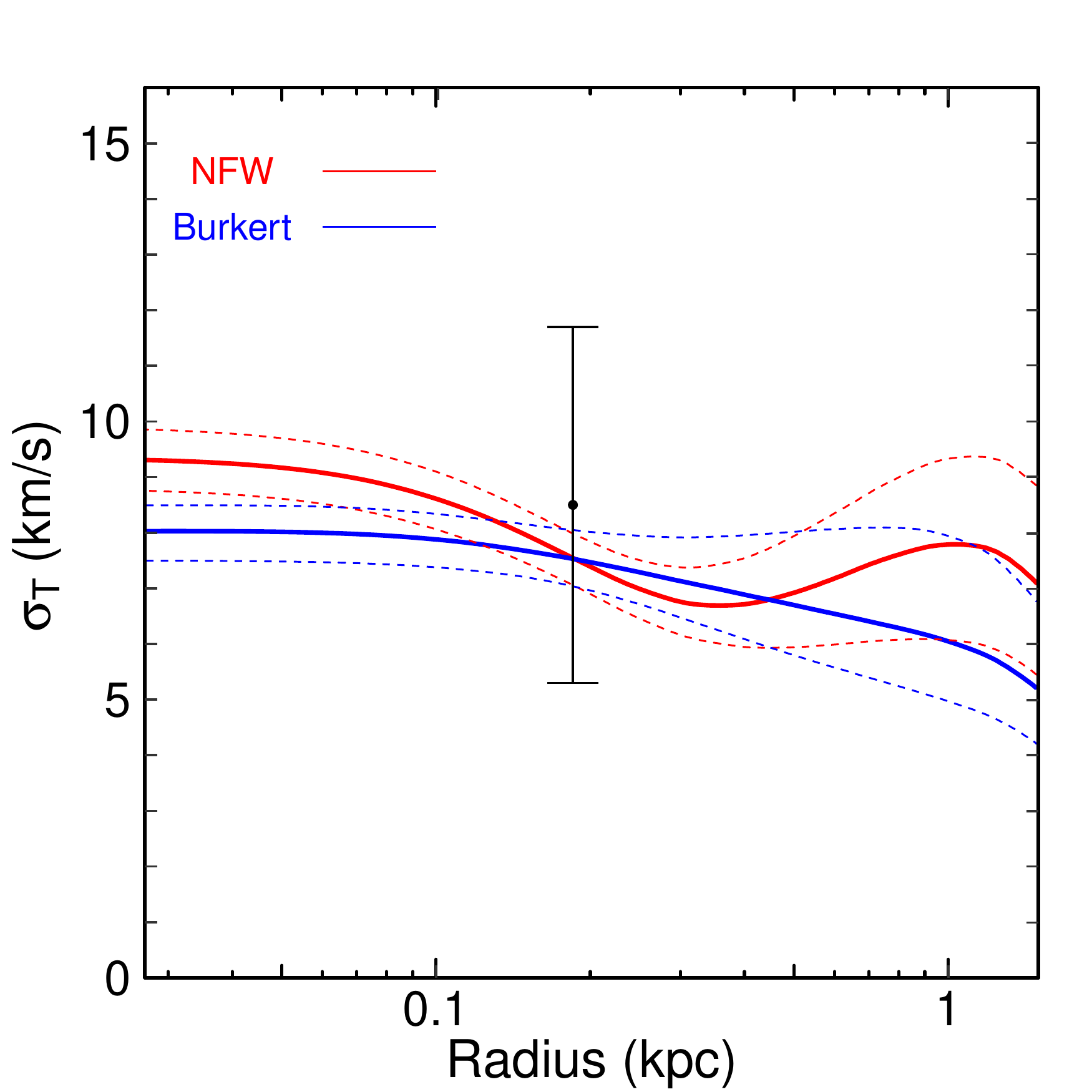}}} & 
{\resizebox{6.0cm}{!}{\includegraphics[angle=0]{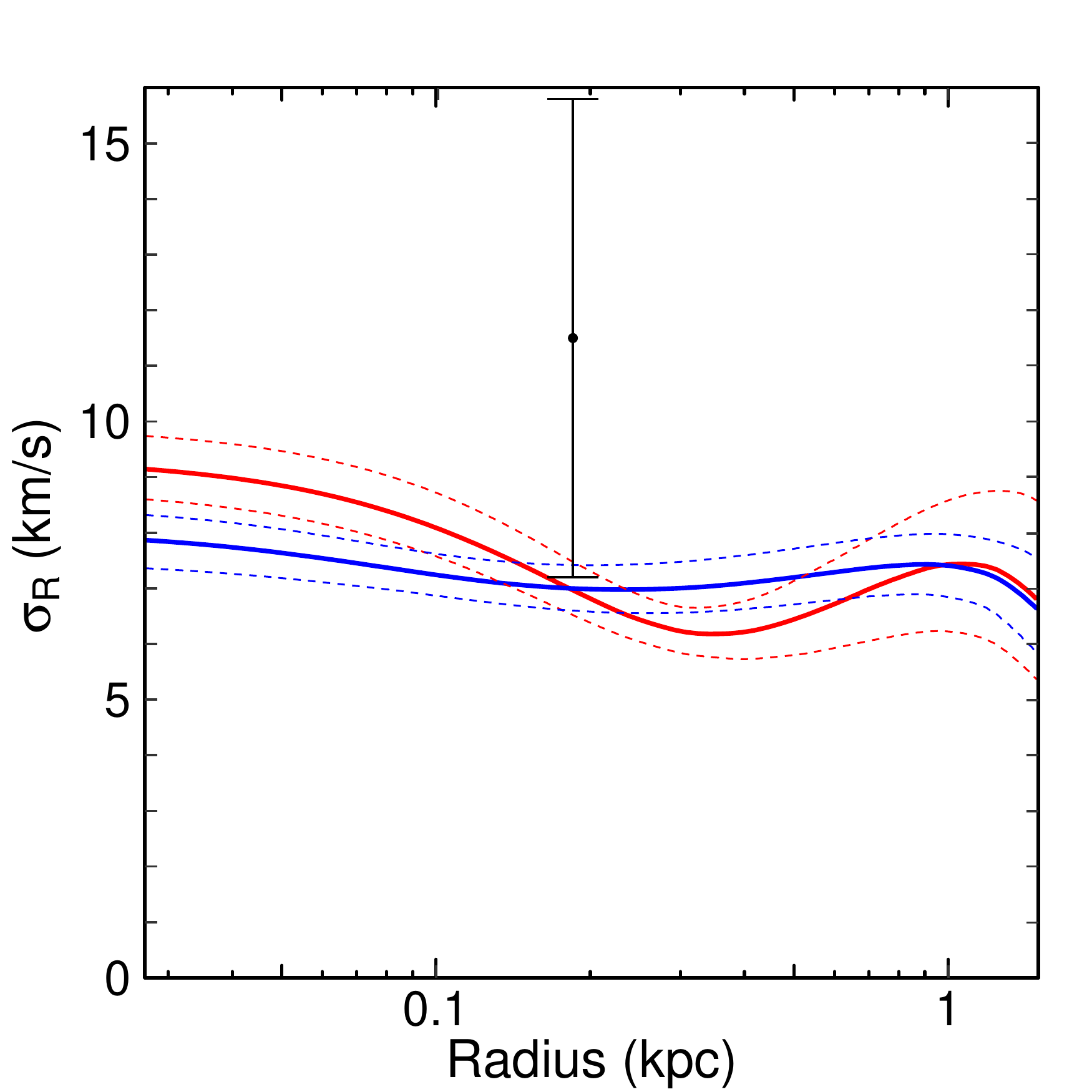}}} &
{\resizebox{6.0cm}{!}{\includegraphics[angle=0]{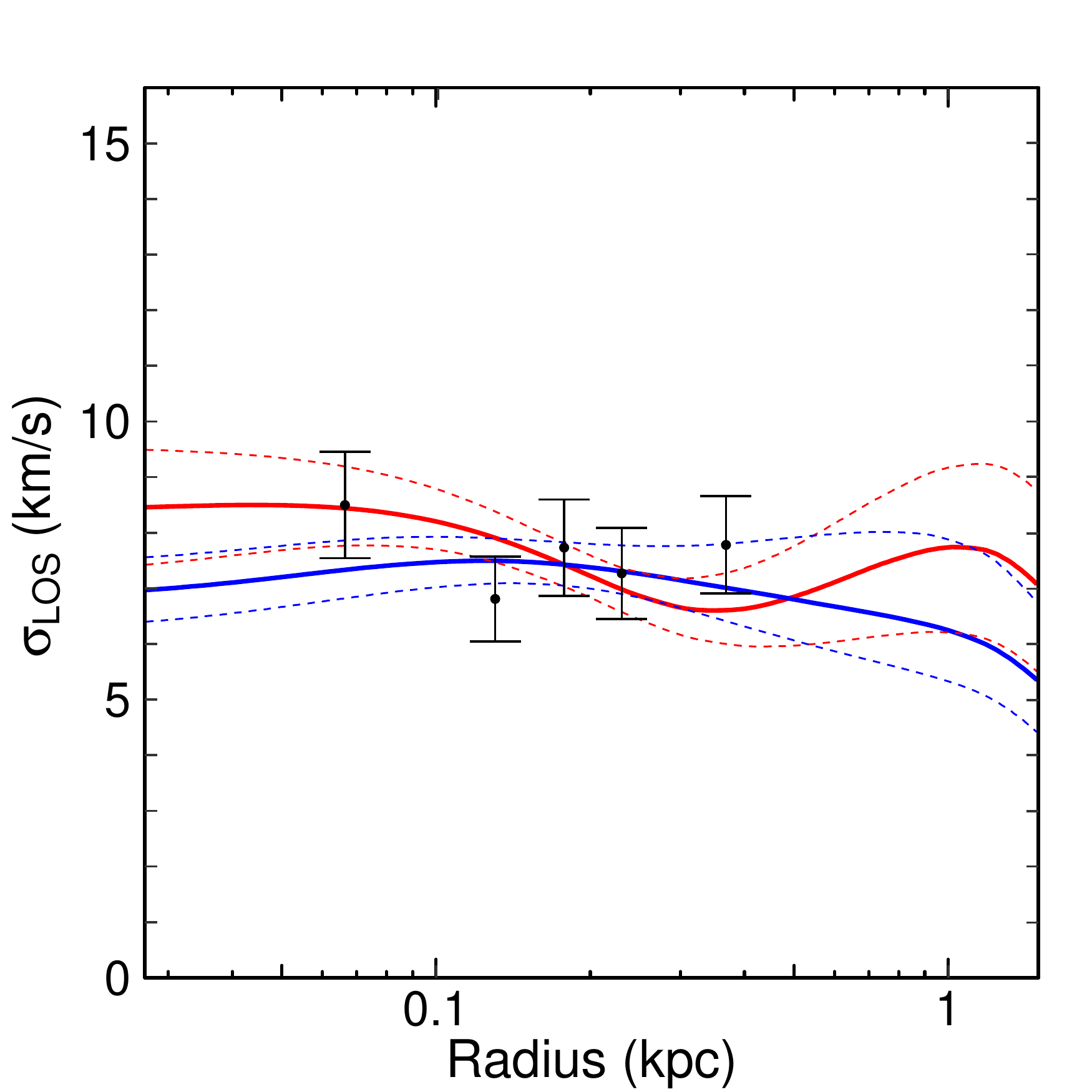}}} \\
{\resizebox{6.0cm}{!}{\includegraphics[angle=0]{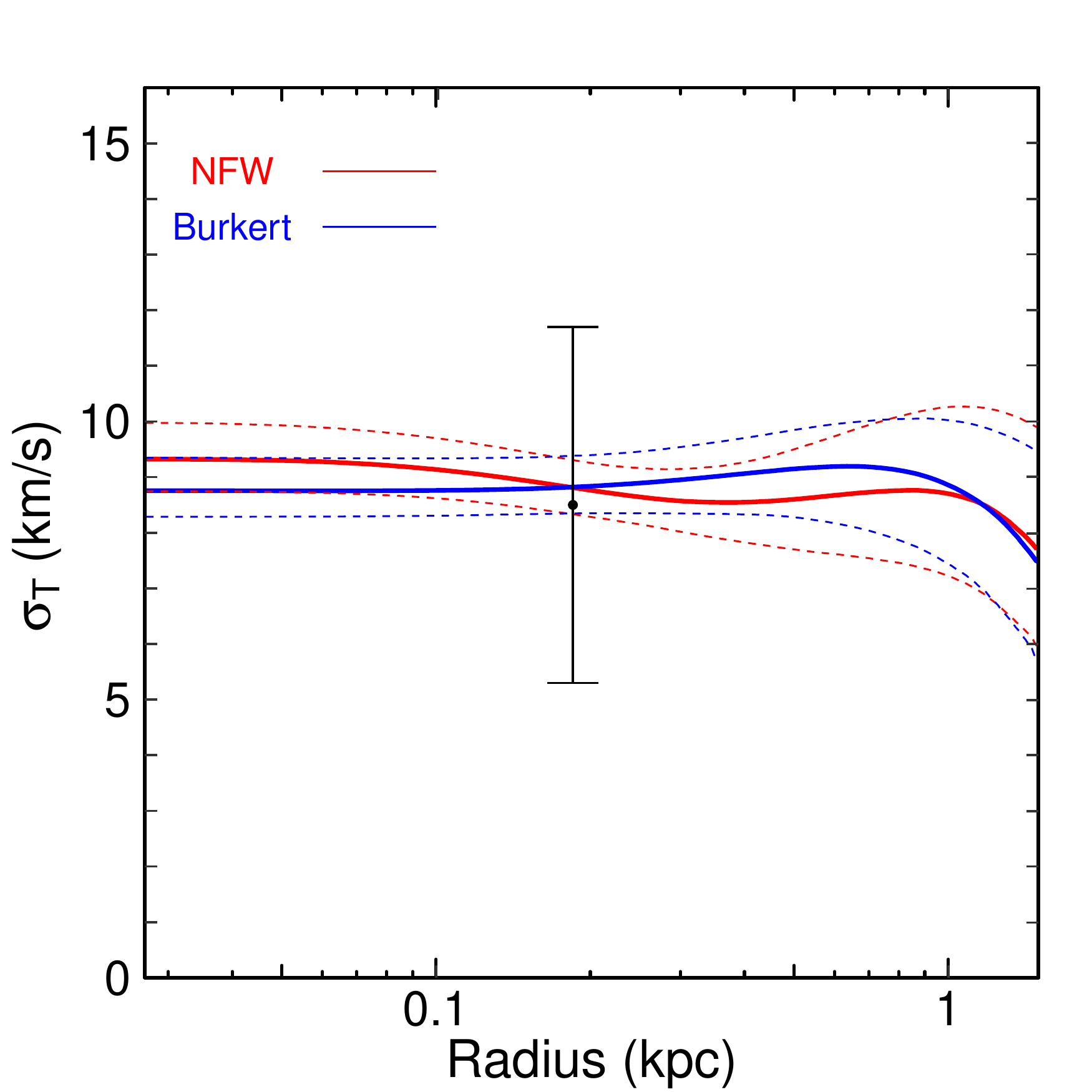}}} & 
{\resizebox{6.0cm}{!}{\includegraphics[angle=0]{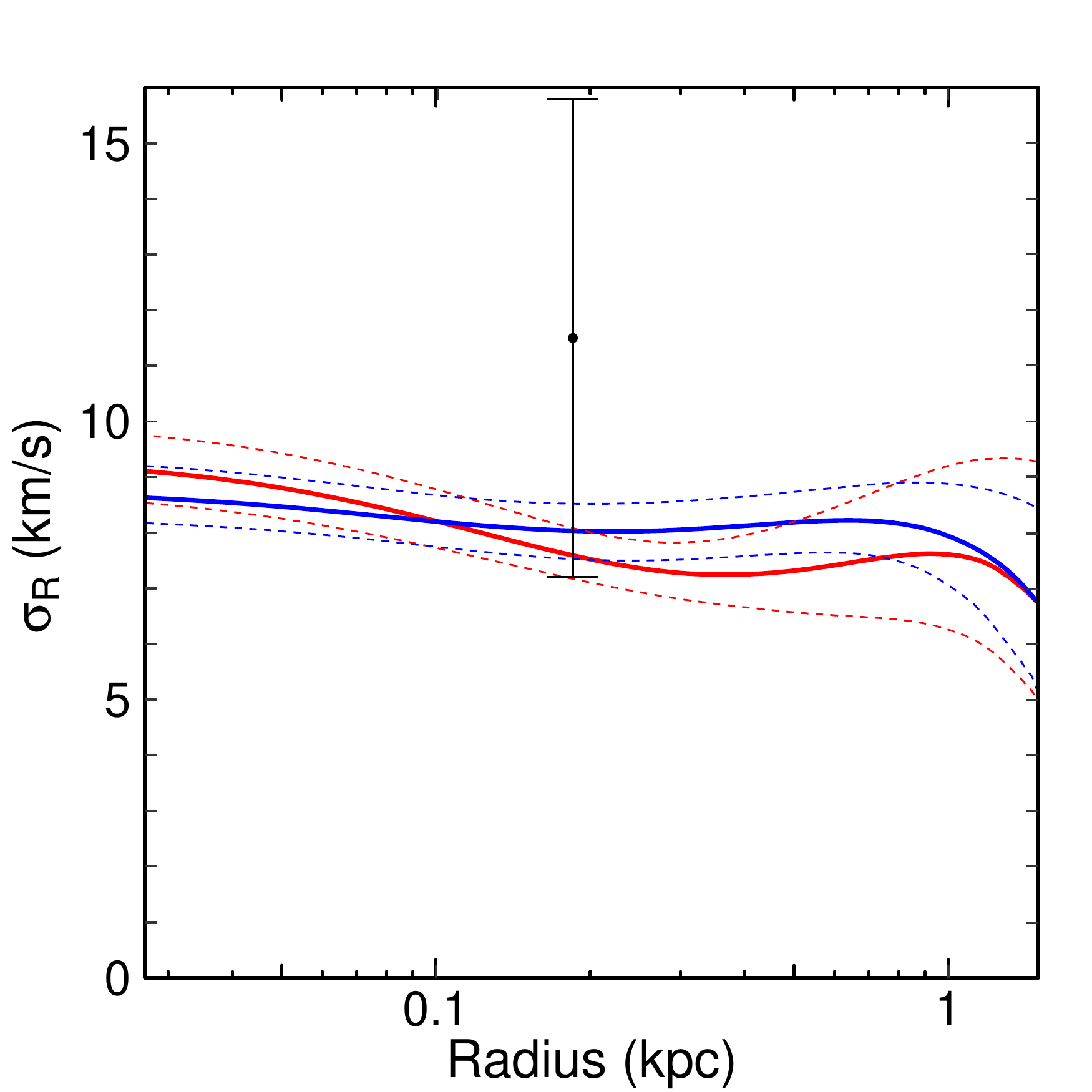}}} &
{\resizebox{6.0cm}{!}{\includegraphics[angle=0]{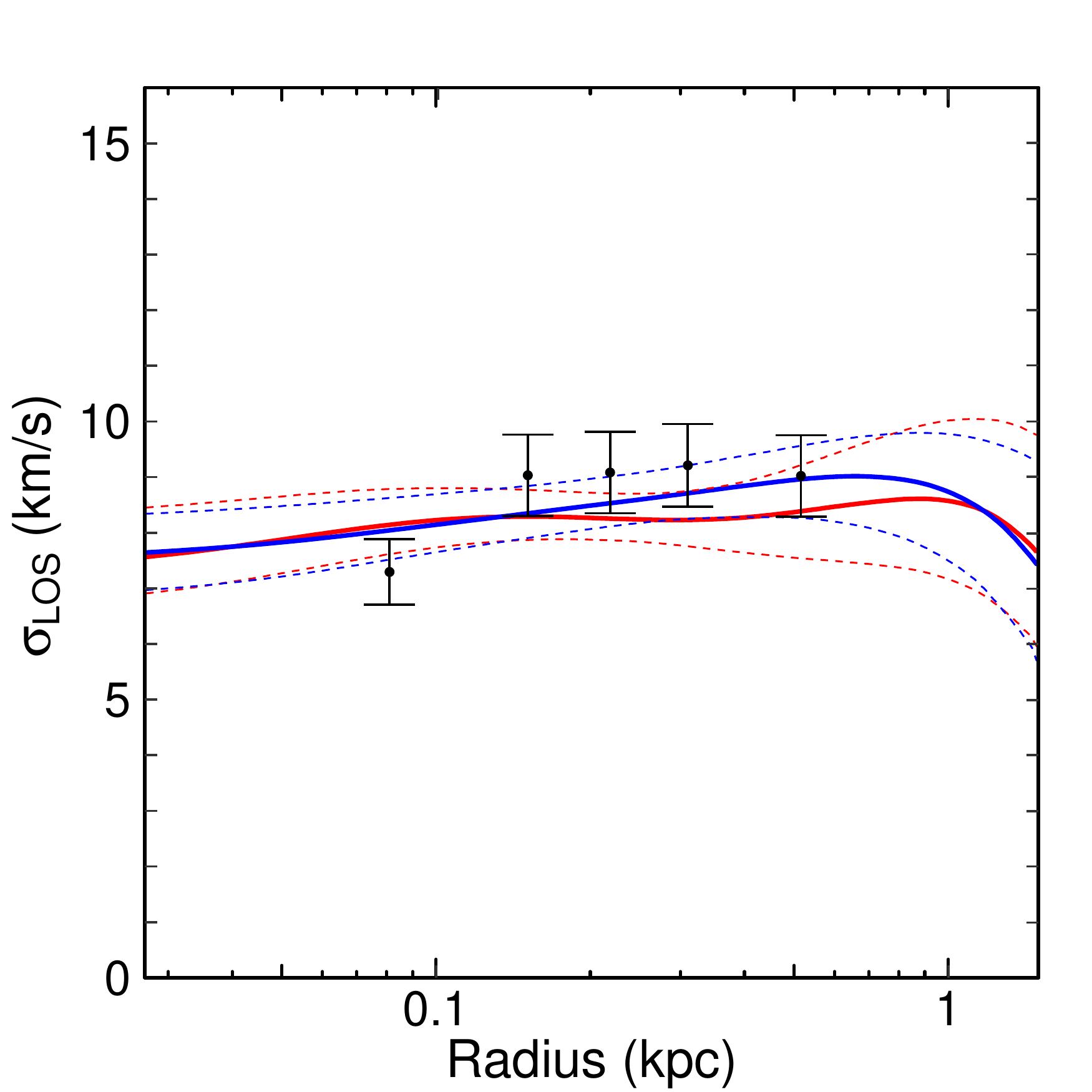}}} \\
{\resizebox{6.0cm}{!}{\includegraphics[angle=0]{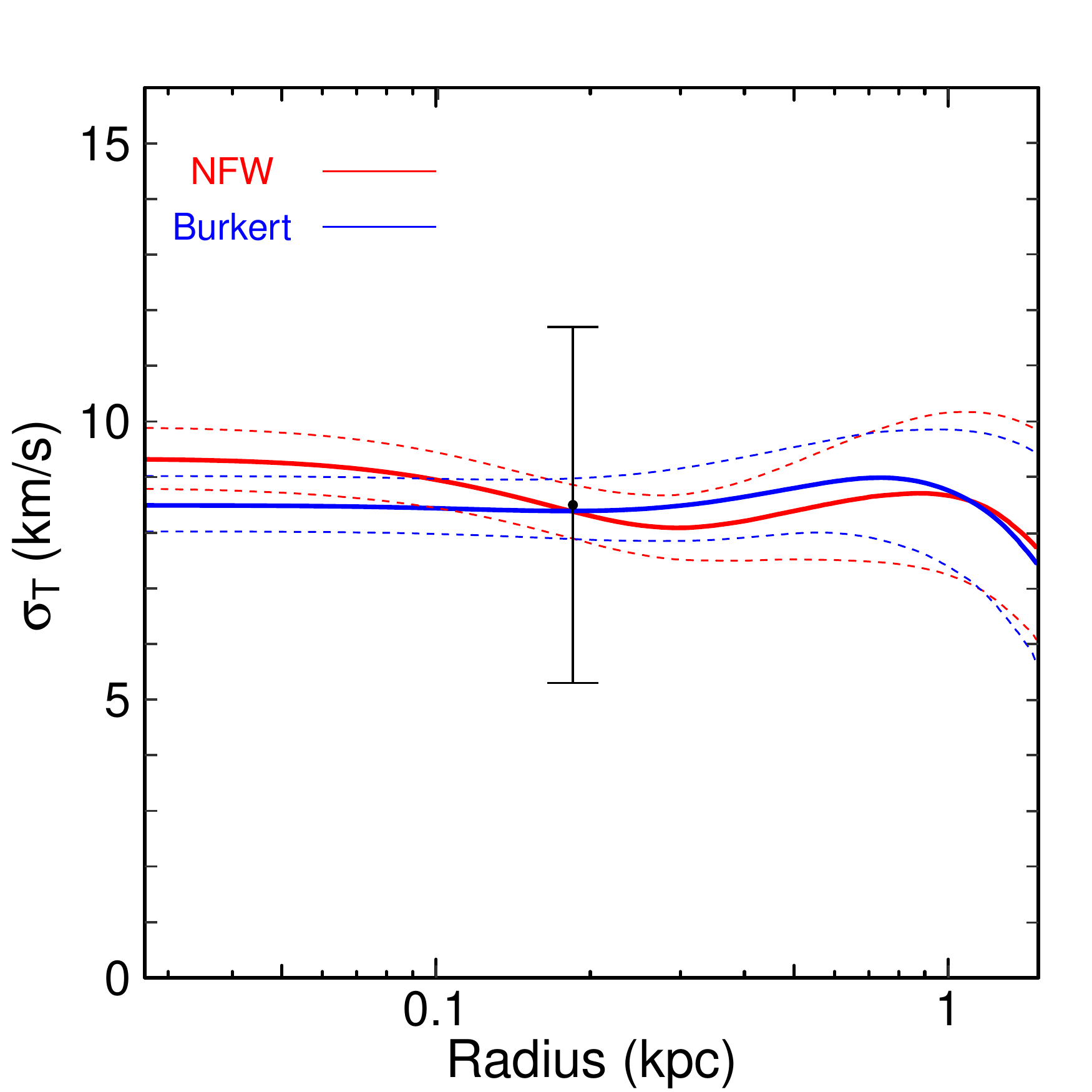}}} & 
{\resizebox{6.0cm}{!}{\includegraphics[angle=0]{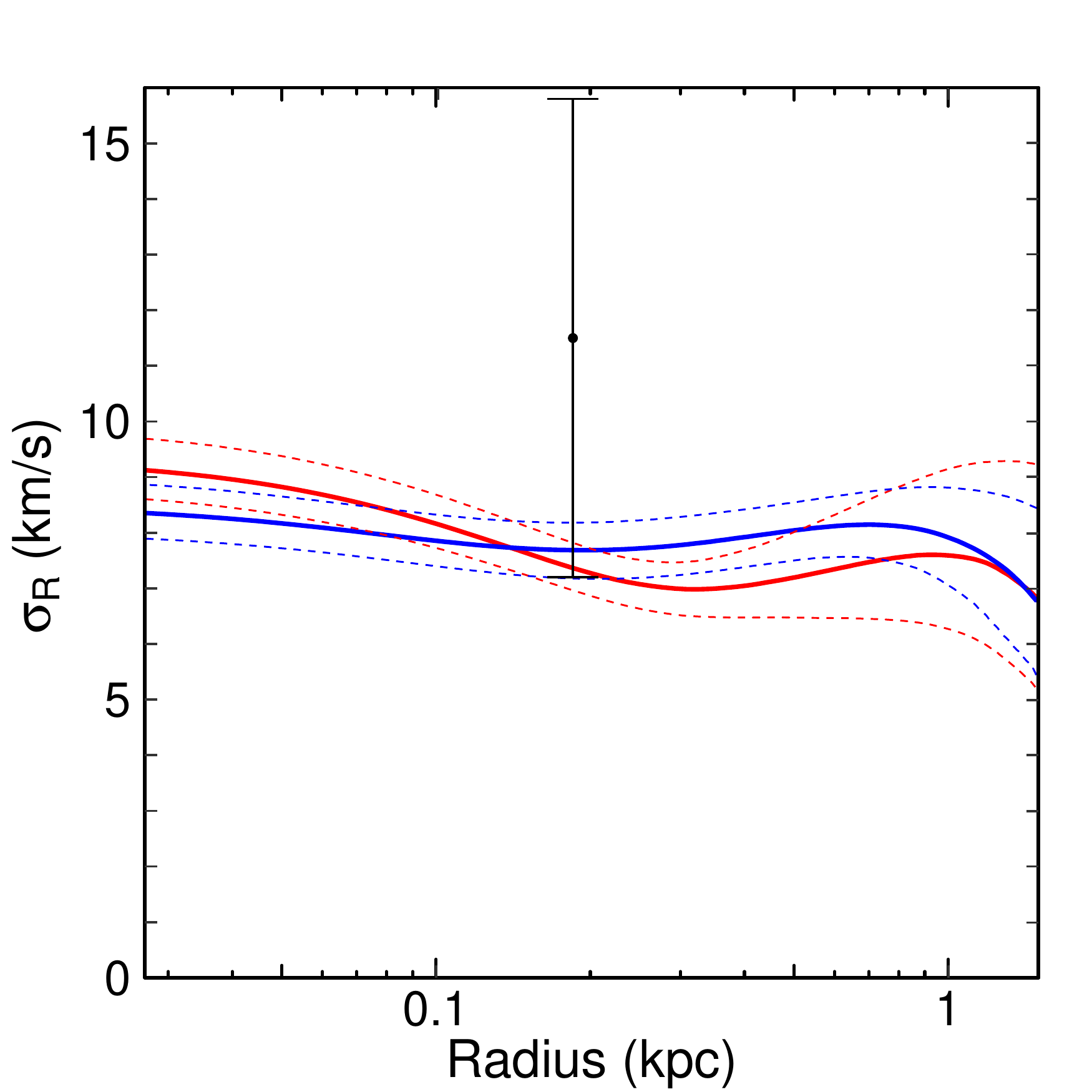}}} &
{\resizebox{6.0cm}{!}{\includegraphics[angle=0]{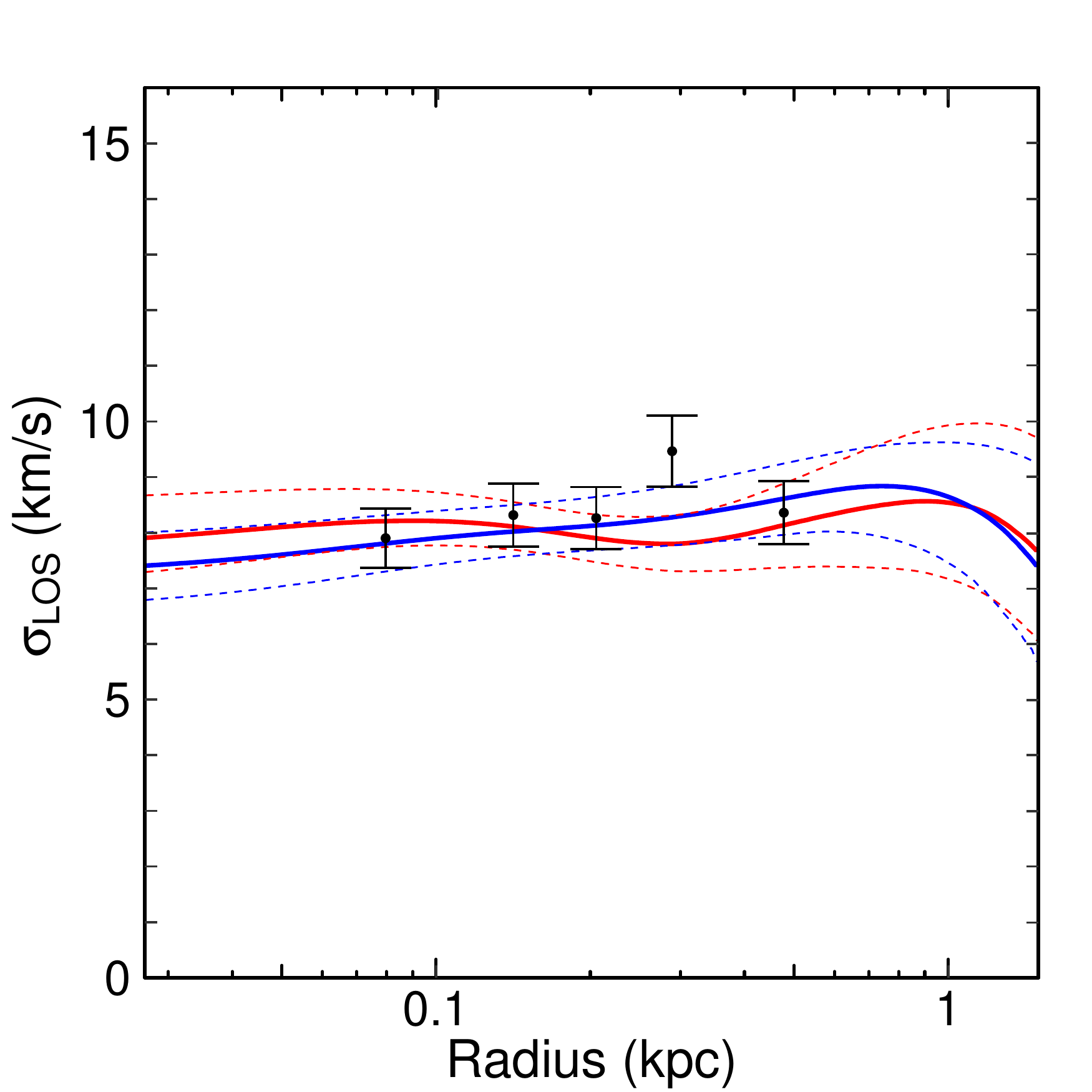}}} \\
\end{tabular}
\end{center}
\caption{Predicted velocity dispersions for the MR (top), MP (middle)
  and combined populations. In each panel, the red is for the NFW
  profile and the blue for the Burkert profile. The solid curves are
  the mean of the posterior distribution at each radius and the dashed
  curves encompass its 10 to 90\% range.  The symbols with error bars
  show the values of the transverse tangential and radial dispersions
  derived from the PM measurements by \citet{2017arXiv171108945M} and
  the LOS dispersion of each population estimated by
  \citet{Strigari:2014yea} based on data from \citet{Walker:2011zu}.}
\label{fig:contours}
\end{figure*}

 While the current PM data are unable to distinguish between the
NFW and the Burkert density profiles, it is interesting to ask whether
this will be possible with a larger sample of PM measurements. A rough
answer to this question is given by Figure~\ref{fig:contours}
where we plot the range of transverse radial, transverse tangential
and line-of-sight velocity dispersion profiles that are compatible with
the observed LOS velocity dispersion and number count data
according to our model. At each projected radius, we calculate the
posterior distribution of each of the three velocity dispersion
components from our Monte Carlo sampling. We show its mean as a solid
curve and enclose its 10 to 90\% range with dashed curves. The points
with error bars in the left and middle panels are the PM-based
measurements of \citet{2017arXiv171108945M}, while for the LOS
velocity dispersion we plot the binned estimates of
\citet{Strigari:2014yea} for the MR, MP and total MR+MP populations.
These are based on the observational data of~\citet{Walker:2011zu}.

The largest difference between the predictions for the NFW and Burkert
dark matter profiles occurs at radii less than the approximate
half-light radius and is most apparent in the MR population. Even in
this case, however, the differences between the two models are less
than 2~km/s. Interestingly, the LOS data alone have very similar
discriminating power as the projected radial and tangential
dispersions.  Taking a simple estimate of the sampling error on each
velocity dispersion as $\sim \sigma_R/\sqrt{2N}$, our results imply
that distinguishing between the NFW and Burkert profiles at the $\sim
2\sigma$ level will require measurement of PMs or LOS velocities for a
few thousand stars with individual uncertainties well below 5~km/s and
spread evenly over the radial range $1.5 < \log_{10} (R/{\mathrm
  1~pc}) < 2.7$.

\section{Discussion and conclusion} 

\par We have performed a dynamical analysis of internal stellar proper
motions in the Sculptor dSph. Using our earlier dynamical model
\citep{Strigari:2014yea}, we compared the predicted tangential
velocity dispersions derived from LOS velocity dispersion and star
count data to direct estimates derived from the recently measured
proper motions of 15 stars by ~\citet{2017arXiv171108945M}. We have
shown that there is excellent consistency between the predicted and
directly measured projected tangential velocity dispersion, while the
predicted and directly measured projected radial velocity dispersions
differ only at $\sim 1 \sigma$.

\par The agreement between the model predictions and the data is just as
good for models that assume an NFW profile for the dark matter as for
models that assume a Burkert profile. Thus, even with the new PM data
we are unable to resolve the long-standing ``core/cusp'' controversy
over the dark matter distribution in the center of Sculptor. However,
we have shown that, in principle, the controversy can be resolved with
larger samples of precise velocities, either line-of-sight or from
PMs, provided a substantial fraction of the measurements are at small
projected radii.  We have checked that splitting the stars into
distinct metallicity populations with different radial distributions
enhances the accuracy with which our distribution function-based
dynamical model can constrain the potential. Thus, obtaining good
quality spectra to allow metallicity determinations is also
important. Finally, improved photometry to reduce uncertainties in the
projected stellar count profiles would also be helpful.

\par The pioneering work of \citet{2017arXiv171108945M} has
demonstrated that internal PM measurements for stars in dSphs can be
made with sufficiently small errors to allow interesting dynamical
modelling. Upcoming {\it Gaia} data releases will lead to a dramatic
improvement in the amount and quality of PM data for bright
dSphs. Together with new ground-based photometric and spectroscopic
campaigns on these objects, this offers the exciting prospect of
finally allowing robust inferences about the central dark matter
distribution in these galaxies and thus about the nature of the dark
matter.

\section{Acknowledgements}
LES is supported by the U.S.~Department of Energy award
de-sc0010813. This work was also supported by the Science and
Technology Facilities Council grants ST/L00075X/1 and ST/P000451/1.
This work used the DiRAC Data Centric system at Durham University,
operated by the Institute for Computational Cosmology on behalf of the
STFC DiRAC HPC Facility (www.dirac.ac.uk). This equipment was funded
by BIS National E-infrastructure capital grant ST/K00042X/1, STFC
capital grant ST/H008519/1, STFC DiRAC Operations grant ST/K003267/1
and Durham University.  DiRAC is part of the National
E-Infrastructure.

\bibstyle{app} 
%\bibliography{bib.bib}

\end{document}